\documentclass[amsmath,amssymb,aps,letterpaper,prl,twocolumn,longbibliography]{revtex4-1}

\usepackage{graphicx}
\usepackage{color}
\usepackage{enumitem}

\usepackage{picture}
\usepackage{calc}

\usepackage{subfigure}
\usepackage{dsfont}
\usepackage{color}

\usepackage[utf8x]{inputenc}

\newcommand{\sfrac}[2]{{\textstyle\frac{#1}{#2}}}
\newcommand{\half}{\sfrac{1}{2}}

\DeclareMathOperator{\tr}{tr}

\newcommand{\lH}{\mathcal{H}}

\newcommand{\lB}{\mathcal{B}}

\newcommand{\bQ}{\mathbb{Q}}

\begin{document}

\title{Classifying nearest-neighbour interactions and deformations of AdS}

\author{Marius de Leeuw}%
 \email{mdeleeuw@maths.tcd.ie}
\author{Chiara Paletta}%
 \email{palettac@maths.tcd.ie}
\author{Anton Pribytok}%
 \email{apribytok@maths.tcd.ie}
\author{Ana.\!\! L.\! Retore}%
 \email{retorea@maths.tcd.ie}
\author{Paul Ryan}%
 \email{pryan@maths.tcd.ie}

\affiliation{%
School of Mathematics
\& Hamilton Mathematics Institute\\
Trinity College Dublin\\
Dublin, Ireland
}%

\begin{abstract}
We classify all regular solutions of the Yang-Baxter equation of eight-vertex type. Regular solutions correspond to spin chains with nearest-neighbour interactions. We find a total of four independent solutions. Two are related to the usual six- and eight-vertex models that have $R$-matrices of difference form. We find two completely new solutions of the Yang-Baxter equation, which are manifestly of non-difference form. These new solutions contain the S-matrices of the $\mathrm{AdS}_2$ and $\mathrm{AdS}_3$ integrable models as a special case. Consequently, we can classify all possible integrable deformations of eight-vertex type of these holographic integrable systems.
\end{abstract}

\maketitle

\section{Introduction}

The Yang-Baxter equation is an important equation that appears in many different areas of physics \cite{doi:10.1142/S0217751X89001503,jimbo1990yang,perk2006yang,Batchelor:2015osa}. It signals the presence of integrable structures which manifest themselves in areas ranging from condensed matter physics to holography. Famous integrable models such as the Heisenberg spin chain and the Hubbard model were important for our understanding of low-dimensional statistical and condensed matter systems.  Similarly, over the last few years, exceptional progress has been made in understanding the AdS/CFT correspondence due to the discovery of integrable structures \cite{Beisert:2010jr}

The solutions of the Yang-Baxter equation are the so-called $R$-matrices which generate the tower of conserved charges that define integrable models \cite{Takhtajan:1979iv,Reshetikhin:1984ug,Sklyanin:1987bi}. Alternatively, they describe two-particle scattering matrices in integrable field theories \cite{ZAMOLODCHIKOV1979253,Zamolodchikov:1990bu}.

Understanding and classifying the solutions of the Yang-Baxter equation is an important and open question with multidisciplinary applications. Recently we put forward a new method  \cite{deLeeuw:2019zsi,deLeeuw:2019vdb} to classify regular solutions of the Yang-Baxter equation by using the so-called boost automorphism \cite{Fuchssteiner1983,Tetelman, HbbBoost,Loebbert:2016cdm}. Regular solutions are those whose corresponding integrable lattice models have nearest-neighbour interactions. 
The main idea behind this method is to use the Hamiltonian rather than the corresponding $R$-matrix as a starting point. So far, we applied this method to solutions of the Yang-Baxter equation that were of difference form $R(u,v) = R(u-v)$. In this paper we extend our approach to the most general case. 

We demonstrate our method by classifying all solutions of the Yang-Baxter equation of eight-vertex type. We find four different types of models. Two models are related to the usual six- and eight-vertex models that have $R$-matrices of difference form. However, additionally, we find two completely new solutions of the Yang-Baxter equation which are manifestly of non-difference form. 

As a further application of our results, we show that the relevant $R$-matrices that appear in the lower-dimensional cases of the AdS/CFT correspondence  \cite{Sfondrini:2014via,Borsato:2014hja,Hoare:2014kma, Hoare:2015kla} are indeed contained in our solutions. We can then use our results to classify all their integrable deformations within the aforementioned framework. We show that the $\mathrm{AdS}_2$ integrable model only admits a one-parameter deformation, while the $\mathrm{AdS}_3$ case admits both a two-parameter elliptic deformation and a family of functional deformations. We postpone further details to an upcoming publication \cite{new}.

\section{Method}
\paragraph{Conserved charges}
Consider a general solution $R(u,v)$ of the Yang-Baxter equation
\begin{align}\label{eq:Yang-Baxter equation}
R_{12} R_{13}R_{23}= 
R_{23} R_{13} R_{12} ,
\end{align}
where we do not assume that $R$ is of difference form \textit{i.e.} $R_{ij}(u_i,u_j)\neq R_{ij}(u_i-u_j)$. Such an $R$-matrix will generate a transfer matrix corresponding to an integrable spin chain via
\begin{align}
t(u,\theta) = \tr_0 \Big[ R_{0L}(u,\theta_L) \ldots R_{01}(u,\theta_1) \Big]   ,
\end{align}
where $L$ would be the number of sites and $\theta_i$ are physical parameters associated to the quantum spaces. We restrict to homogeneous spin chains in which the $\theta_i = \theta$ parameters of all  physical spaces coincide.

We furthermore restrict to integrable models with nearest-neighbour interactions and hence we assume that $R$ is regular, \textit{i.e} $R_{ij}(u,u) =P_{ij}$ where $P_{ij}$ is the permutation operator on sites $i$ and $j$. Then, the spin chain Hamiltonian $\bQ_2$ has interaction range two and is given by the logarithmic derivative of the transfer matrix
\begin{align}\label{eq:hamiltonian} 
&\bQ_2(\theta) = \sum_i \lH_{i,i+1}(\theta),
&&\lH(\theta) = P \frac{d R(u,\theta)}{d u} \Big |_{u=\theta}.
\end{align}
%
%
%
In the special case when the $R$-matrix is of difference form, the dependence on the parameter $\theta$ drops out. 

The other conserved charges of the integrable model are given by the higher derivatives of the transfer matrix. More specifically we have
\begin{align}
\bQ_{r+1} \sim \frac{d^r}{du^r} \log t(u,\theta) \Big|_{u=\theta}.
\end{align}
The interaction range of $\bQ_r$ is $r$ and from \eqref{eq:Yang-Baxter equation} it follows that
\begin{align}
[\bQ_r,\bQ_s]=0.
\end{align}
This tower of conserved charges is the defining property of an integrable system. In this paper we will construct all models with certain properties that have a tower of conserved charges coming from an $R$-matrix.

\paragraph{Boost operator}
Instead of taking derivatives of the transfer matrix, there is an alternative way to compute the higher conserved charges $\bQ_{r=3,4,.\ldots}$. Namely, the so-called boost operator $\lB[\bQ_2]$ satisfies \cite{Tetelman, HbbBoost,Loebbert:2016cdm}
\begin{align}
&\bQ_{r+1} \sim [\lB[\bQ_2],\bQ_r],
&& r>1.
\end{align}
The boost operator is a differential operator and depends on the coefficients of the Hamiltonian \cite{HbbBoost}
\begin{equation}\label{eq:boost}
\lB[\bQ_2]:=\partial_\theta +  \sum_{n=-\infty}^\infty n\, \lH_{n,n+1} (\theta).
\end{equation}
This expression is strictly-speaking only defined for infinite length chains, but reduces consistently to spin chains of finite length. 

\paragraph{Integrable Hamiltonians}
Now we consider a nearest-neighbour Hamiltonian with general entries $h_{ij}(\theta)$ and compute the corresponding charge $\bQ_3$ by using the boost operator \eqref{eq:boost}. The Hamiltonian potentially corresponds to an integrable system if 
\begin{align}\label{eq:IntCond}
[\bQ_2,\bQ_3] = 0.
\end{align}
This is a necessary condition for integrability and it takes the form of a set of coupled first order, non-linear, differential equations for the components of $\lH$.

\paragraph{$R$-matrix}
In order to prove integrability we then, for each potentially integrable Hamiltonian, compute the corresponding $R$-matrix. 
 Let $\dot{R}$ be the derivative with respect to the first variable, then by expanding the Yang-Baxter equation around the point $u_1=u_2\equiv u$ to first order we find
\begin{align}\label{eqn:Yang-Baxter equationfirstorder1}
\Big[R_{13} R_{23}, \lH_{12}(u)\Big]  &= \dot{R}_{13} R_{23} - R_{13} \dot{R}_{23}.
\end{align}
Similarly with $ R^\prime $ denoting the derivative with respect to the second variable,  expanding the Yang-Baxter equation around $ u_2=u_3\equiv v$ yields
\begin{align}\label{eqn:Yang-Baxter equationfirstorder2}
	\Big[R_{13} R_{12}, \lH_{23}(v)\Big]  &=  R_{13}R^\prime_{12} - R^\prime_{13}R_{12}  ,
\end{align}
with $R_{ij} = R_{ij}(u,v)$. These equations are special cases of the Sutherland equation \cite{Sutherland}
and they form a set of coupled first order differential equations. Since we assume regularity and know the Hamiltonian, we see that we obtain two boundary conditions which in principle fix our solution uniquely. Subsequently, we can verify whether the solutions of the Sutherland equations satisfy the Yang-Baxter equation and formally prove integrability. Notice that this method is complete in the sense that any solution of the Yang-Baxter equation necessarily gives an integrable Hamiltonian.

\section{Identifications}\label{sec:Iden}

There are some simple ways in which different solutions of the Yang-Baxter equation can be related to each other. In what follows we will identify models which can be mapped to each other under the following transformations.

\paragraph{Local basis transformation} 
If $R(u,v)$ is a solution of the Yang-Baxter equation, then we can generate a different regular solution by defining
\begin{align}
R^{(V)}(u,v) = \Big[V(u)\otimes V(v)\Big] R(u,v)  \Big[V(u)\otimes V(v)\Big]^{-1}.
\end{align}
%
It gives rise to a new integrable Hamiltonian 
\begin{align}
\lH^{(V)} = \big[\!V\otimes V\big] \lH  \big[\!V\otimes V\big]^{-1}\! - \big[\dot{V} V^{-1}\otimes 1 - 1 \otimes \dot{V} V^{-1}\big],
\end{align}
where everything is evaluated at $\theta$.

\paragraph{Reparametrization}

If $R(u,v)$ is a solution, then $R(f(u),f(v))$  is a solution of the Yang-Baxter equation as well. This transformation affects the normalization of the Hamiltonian. We are also free to reparameterize any other functions and constants in both the $R$-matrix and Hamiltonian. For instance the $R$-matrices from \cite{Stouten:2017azy,Stouten:2018lkr} can be obtained by  a reparameterization of the  XXX $R$-matrix.

\paragraph{Normalization}

If $R(u,v)$ is a solution, then for any function $g(u,v) $ the product $g(u,v) R(u,v)$ is also a solution. On the level of the Hamiltonian this corresponds to rescaling and shifting $g(\theta,\theta)\lH+g'(\theta,\theta)$.

\paragraph{Discrete transformations} For any solution $R(u,v)$ of the Yang-Baxter equation, $PR(u,v)P, R^T(u,v)$ and $PR^T(u,v)P$ are solutions as well.

All these transformations are universal and hold for any integrable model. They have a trivial effect on the spectrum, which means that they describe the same physical model. 

\paragraph{Twists} Additionally, there are identifications that are model dependent. 
 If $R$ is a solution and assuming $[U(u)\otimes U(v),R]=[V(u)\otimes V(v),R]=0$, then $(U(u)\otimes V(v)) R (V(u)\otimes U(v))^{-1}$ is also a solution.
 A twist will generically have a non-trivial effect on the spectrum of the model.

\section{Results for $4\times 4$}

We apply our method to spin chains with a two-dimensional local Hilbert space. We already applied our method to such $R$-matrices of difference form and found that only Hamiltonians of eight-(or less) vertex type seem to be physical \cite{deLeeuw:2019zsi}. For this reason we will for the moment only consider Hamiltonians of this type. We parametrize our Hamiltonian as
\begin{align}
\begin{split}
\mathcal{H}  =\,& 
h_1  \text{ }\mathds{1} + h_2 (\sigma _z\otimes \mathds{1}- \mathds{1}\otimes \sigma _z) + h_3  \sigma _+\otimes \sigma _-  + \\
& h_4 \sigma_-\otimes \sigma _+ +    h_5
( \sigma _z \otimes \mathds{1} +  \mathds{1} \otimes \sigma _z ) + \\
&h_6 \sigma _z\otimes \sigma _z  +
 h_7 \sigma _-\otimes
\sigma _- + h_8 \sigma _+\otimes \sigma _+,
\end{split}
\label{generalHamiltonian}
\end{align}
where $ h_i=h_i(\theta) $ and $ \sigma_i $ are the Pauli matrices with $ \sigma^\pm=\half(\sigma_x\pm i \sigma_y) $.
Let us also introduce the primitive functions $H_i(\theta) = \int^\theta h_i(\phi)\mathrm{d}\phi$. 

Similarly we write our $R$-matrix as
\begin{align}
\begin{split}
R  =\begin{pmatrix}
r_1 & 0 & 0 & r_8 \\
0 & r_2 & r_6 & 0 \\
0 & r_5 & r_3 & 0 \\
r_7 & 0 & 0 & r_4 \\
\end{pmatrix}
\end{split},
\label{generalR}
\end{align}
where we suppressed the dependence on the spectral parameters.

After our identifications, we find only four independent types of $4\times 4$ Hamiltonians that solve the integrability condition \eqref{eq:IntCond}
\begin{itemize}[noitemsep]
\item 6-vertex A, $h_6 \neq 0$ and $h_7=h_8 = 0$ 
\item 6-vertex B, $h_6 = h_7=h_8 = 0$ 
\item 8-vertex A, $h_6\neq0,h_7\neq0,h_8\neq0$
\item 8-vertex B, $h_6=0$ and $h_7\neq0,h_8\neq0$.
\end{itemize}
Notice that for $R$-matrices of difference form, there are  eight independent solutions \cite{Vieira:2017vnw,deLeeuw:2019zsi}. This means that some of these solutions are reductions of the same non-difference $R$-matrix. For example, all 7-vertex type solutions are special cases of 8-vertex models. If a Hamiltonian is equivalent to a well-known one under identifications we will not list its $R$-matrix.

Let us discuss these models in more detail.

\paragraph{6-vertex A} 
Setting $h_7=h_8=0$ and plugging this Hamiltonian into \eqref{eq:IntCond}, we see that it is satisfied if and only if the functions $h_i(\theta)$ satisfy the following differential equations
\begin{align}
&\frac{\dot{h}_3}{h_3} =  \frac{\dot{h}_6}{h_6}+4h_5,
&&\frac{\dot{h}_4}{h_4} = \frac{\dot{h}_6}{h_6}-4h_5,
\end{align}
provided that $h_6\neq0$. This is easily solved to give
\begin{align}
&h_3 = c_3 h_6 e^{4 H_5},
&&h_4 = c_4 h_6 e^{-4 H_5},
\end{align}
where $c_{3,4}$ are constants. The Hamiltonian is equivalent to that of the XXZ spin chain. In other words, the source of the non-difference dependence on the spectral parameters is only due to twists, basis transformations and reparameterizations.

\paragraph{6-vertex B} It is easy to see that setting $h_6=h_7=h_8=0$ makes the Hamiltonian satisfy $[\bQ_2,\bQ_3]=0$ for \textit{any} choice of $h_1, \ldots, h_5$. Thus, the Hamiltonian depends on five free functions. We can account for four of them by using a local basis transformation, a twist, a normalization and a reparameterization. Since there is one free function left, this model does not have an $R$-matrix of difference form underlying it and is a new solution of the Yang-Baxter equation. 

Without loss of generality, we normalize our $R$-matrix such that $r_5=1$ and set $h_2 = 0$. It then follows from the Sutherland equation \eqref{eqn:Yang-Baxter equationfirstorder1} that $r_6=1 = r_1 r_4 + r_2 r_3 $ and 
\begin{align}
&r_1 = \frac{\dot{r}_2 +2 h_5 r_2}{h_4},
&&r_3  =  -\frac{2 h_5 r_4 + \dot{r}_4}{h_4},
\end{align}
while $r_4$ satisfies a Riccati equation
\begin{align}
\ddot r_4-\frac{\dot{h}_4}{h_4}\dot{r}_4+r_4\Big[h_3 h_4-\frac{2 h_5 \dot {h}_4}{h_4}+2(\dot{h}_5-2 h_5^2)\Big]=0.
\end{align}
Now we introduce a reparameterization of the spectral parameter 
\begin{align}
u_i  \mapsto x_i =2 \int^{u_i}  \frac{h_5 \frac{\dot h_4}{h_4}-\dot h_5}{h_3 h_4-4 h_5^2},
\end{align}
which kills the non-derivative term in the Riccati equation. It is then straightforward to solve our system of differential equations to find $r_2(x,y) = H_4(x) - H_4(y)$ and
\begin{align}
r_1(x,y) &= r_4(y,x) = 1 + 2 \frac{h_5(x)}{h_4(x) }r_2(x,y) , \\
r_3(x,y) &= 4  \frac{h_5(x)}{h_4(x) } \frac{h_5(y)}{h_4(y) }r_2 -2\Big[ \frac{h_5(x)}{h_4(x) }- \frac{h_5(y)}{h_4(y)} \Big],
\end{align}
This solution is manifestly of non-difference form and it is easy to show that it satisfies the Yang-Baxter equation and the correct boundary conditions. Notice also that the form of the $R$-matrix depends on the type of functions $h_i$ from the Hamiltonian. For instance, if $h_i$ are constants then $R$ will be rational. 

\paragraph{8-vertex A} In case $h_6\neq0$, the integrability constraint gives that $h_3=h_4$, $h_5 =0$ together with the following equations
\begin{align}
& \frac{\dot{h}_3}{h_3} = \frac{\dot{h}_6}{h_6},
&& \frac{\dot{h}_7}{h_7} = \frac{\dot{h}_6}{h_6}+ 4 h_2,
&& \frac{\dot{h}_8}{h_8} = \frac{\dot{h}_6}{h_6}- 4 h_2,
\end{align}
which are easily solved by
\begin{align}
& h_3 = c_3 h_6,
&& h_7 = c_7 h_6 e^{4H_2},
&& h_8 = c_8 h_6 e^{-4H_2},
\end{align}
where $c_i$ are constants. The resulting Hamiltonian is that of the XYZ spin chain under our identifications.

\paragraph{8-vertex B} In the case when $h_6=0$, we find that the most general solution satisfies the differential equations
\begin{align}
&   \frac{\dot{h}_7}{h_7} = 4 h_2 +  \frac{\dot{h}_3+\dot{h}_4}{h_3+h_4} + 4 \frac{h_3-h_4}{h_3+h_4} h_5 =  \frac{\dot{h}_8}{h_8} + 8h_2,\\
& \frac{\dot{h}_5}{h_5} =- \frac{h_3^2-h_4^2}{4 h_5}+ \frac{\dot{h}_3+\dot{h}_4}{h_3+h_4} +4\frac{h_3-h_4}{h_3+h_4} h_5.
\end{align}
In order to solve these equations we  introduce two new functions that simplify this set of differential equations. Defining $b_1,b_2$ such that
\begin{align}
&h_3 =  \pm\sqrt{\frac{b_1}{b_2}}(2 h_5+b_2),
&& h_4 = \pm\sqrt{\frac{b_1}{b_2}}(2 h_5 - b_2),
\end{align}
we get a simple equation for $b_2$ that can be solved to give
\begin{align}
& b_2 = \frac{b_1 }{c_2^2 e^{4B_1}+1},
&&  B_1 = \int b_1.
\end{align}
The solutions to the remaining equations are then
\begin{align}
&  h_7 =c_7h_5e^{4H_2+2B_1},
&&  h_8 = c_8h_5e^{-4H_2+2B_1}.
\end{align}
We see that there are four free functions remaining and hence this model is genuinely of non-difference form. 

We again normalize our $R$-matrix such that $r_5=1$ and we use a local basis transformation to set $h_2 = 0$. We then apply a further constant basis transformation and set $h_7 = h_8$ which implies that $r_7 = r_8$. Moreover, let us set the normalization of $\lH$ such that $h_8 = k$, which corresponds to choosing $h_5=\frac{k}{c_8}e^{-2B_1}$. We see that $r_5=r_6$ and obtain the following differential equation for $r_8$
\begin{align}
\dot{r}_8^2 =k^2( r_8^2+1)^2-4r_8^2.
\end{align}
This can only be solved in closed form since $k = \frac{c_7}{2c_2}$ is constant. The solution is
\begin{align}
r_8(u,v) = k \frac{
\mathrm{sn} (u-v,k^2) \mathrm{cn}(u-v,k^2)}{\mathrm{dn}(u-v,k^2)},
\end{align}
where $\mathrm{sn,cn,dn}$ are the  Jacobi elliptic functions with modulus $k^2$. The remaining entries of $R$ can be expressed in terms of $r_8$ and after redefining $h_5(x) = -\frac{1}{2}\cot \eta(x)$ we find
\begin{align}
r_1 &= 
\frac{1}{\sqrt{\sin \eta(u)}\sqrt{\sin\eta(v)}}  \bigg[\sin\eta_+\frac{\mathrm{cn}}{\mathrm{dn}} 
-\cos\eta_+ \mathrm{sn}\bigg], \\
r_2 &= 
\frac{\mp 1}{\sqrt{\sin \eta(u)}\sqrt{\sin\eta(v)}}  \bigg[\cos\eta_-\mathrm{sn} +\sin\eta_-\frac{\mathrm{cn}}{\mathrm{dn}} 
\bigg],\\
r_3 &= 
\frac{\mp 1}{\sqrt{\sin \eta(u)}\sqrt{\sin\eta(v)}}  \bigg[\cos\eta_-\mathrm{sn}  - \sin\eta_-\frac{\mathrm{cn}}{\mathrm{dn}} 
\bigg],\\
r_4 &= 
\frac{1}{\sqrt{\sin \eta(u)}\sqrt{\sin\eta(v)}}  \bigg[\sin\eta_+\frac{\mathrm{cn}}{\mathrm{dn}} 
+\cos\eta_+ \mathrm{sn}\bigg],
\end{align}
where $\eta_\pm = \frac{\eta(u) \pm \eta(v)}{2}$ and all the Jacobi elliptic functions depend on $u-v$, \textit{i.e.} $\mathrm{sn} = \mathrm{sn}(u-v,k^2)$.
This solution indeed satisfies the Yang-Baxter equation and has the correct boundary conditions. Moreover, it is easy to see that in the case where $\eta$ is constant, it becomes 
of difference form and reduces to the well-known solution found in \cite{8v,Khachatryan:2012wy,Vieira:2017vnw}.

The limit $c_2\rightarrow 0$ is interesting, and the models falling into this category include the $\mathrm{AdS}_2$ integrable system. However, it should be handled with certain care and is equivalent to take $k\rightarrow \infty$. In order to take this limit we should rescale the spectral parameters $(u,v)\mapsto (\frac{u}{k},\frac{v}{k})$ and make use of the identities for inversions of the elliptic modulus. We also need to redefine $B_1\left(\frac{u}{k}\right)\rightarrow B_1(u)$.
We can then safely take $k\rightarrow \infty$ and find that $R$ becomes of trigonometric type. 
%
\section{Deformations of $\mathrm{AdS}_{2,3}$}

For both the  $\mathrm{AdS}_{2}$ and  $\mathrm{AdS}_{3}$ integrable models, the $R$-matrix contains separate $4\times 4$ blocks that need to satisfy the Yang-Baxter equation by themselves. We demonstrate that these blocks fit into our classification. From our method, we note that it is enough to map the  $\mathrm{AdS}_{2,3}$ Hamiltonians to the Hamiltonians that we found, rather than compare $R$-matrices. The AdS/CFT Hamiltonians depend on the rapidity through the variables $x^\pm$ defined as
\begin{align}
& u = \frac{1}{2}\Big[x^+ + \frac{1}{x^+} +x^- + \frac{1}{x^-}\Big],&& \frac{x^+}{x^-}=e^{i p}.
\end{align}

\paragraph{AdS$_3$} For  $\mathrm{AdS}_{3}$, we see that the Hamiltonian of particles with the same chirality \cite{Sfondrini:2014via,Borsato:2014hja} is of 6-vertex B type. We compute the Hamiltonian and identify the resulting functions with $h_1,\dots,h_5$. For the spin chain frame \cite{Borsato:2013qpa}, the result is given by $h_2=0$ and 
\begin{align}
& h_3(u)=\frac{\dot{x}^-}{x^--x^+}, && h_4(u)=\frac{\dot{x}^+}{x^--x^+}, \\
&h_1 = -\sfrac{1}{2}(h_3+h_4),
&& h_5=-\half h_1
\end{align}
and we take the positive sign in the square root in $h_3$, $h_4$. A similar expression holds for \cite{Borsato:2014hja} up to factors of $e^{ip/2}$ in $h_{3,4}$ and now $h_2\neq 0$.
We now see that there are two possible types of deformations. First, we can match this model with our 6-vertex B type, which leaves us with a continuous family of deformations since we can add arbitrary functions of the spectral parameter to all of the components. This might be a reflection of the special nature of two-dimensional CFTs. Second, we can embed the Hamiltonians  \cite{Borsato:2013qpa,Borsato:2014hja} in our 8-vertex B model. This gives a one-parameter elliptic deformation of the $\mathrm{AdS}_3$ model. The embedding is given, for the spin chain frame, by
\begin{align}
& h_1 = \frac{1}{2}\frac{\dot{x}^+ + \dot{x}^-}{x^+-x^-}, && b_1=\frac{1}{2}\frac{\dot{x}^+-\dot{x}^-}{x^+-x^-},&& h_5=-\half h_1.
\end{align}
Together with $h_2=c_2 = c_7 = c_8 = 0$. This is a novel elliptic deformation. For the string frame, $c_2\neq 0$ and $h_2\neq 0$ and we take the positive sign of the square root in $h_{3,4}$.

\paragraph{AdS$_2$} The massive sector of the $\mathrm{AdS}_2\times S^2 \times T^6$ string sigma model \cite{Hoare:2014kma, Hoare:2015kla} is of  8-vertex B  type with the $+$ sign in the square root in the Hamiltonian. It has $c_2=0$ and furthermore $c_7=-c_8$. The non-zero components of the Hamiltonian are parameterized as
\begin{align}
& h_1 = \frac{1}{4}\frac{x^-+x^+}{x^--x^+} \Big[\, \frac{\dot{x}^+}{x^+}+\frac{\dot{x}^-}{x^-}\Big],\\
& h_5 = \frac{1}{8}\frac{1+e^{\frac{ip}{2}}}{1-e^{\frac{ip}{2}}} \Big[\, \frac{\dot{x}^+}{x^+}+\frac{\dot{x}^-}{x^-}\Big],\\
& B_1 = -\frac{1}{2}\log\left[\frac{c_8 e^{-i\frac{p}{2}} }{4} \frac{1+e^{\frac{ip}{2}}}{1-e^{\frac{ip}{2}}}\Big(x^+-\frac{1}{x^-}\Big)\right].
\end{align}
We conclude that this integrable model only admits a one-parameter deformation by taking $c_2$ to be non-zero.

\section{Conclusions \& Outlook}

In this paper we classified all regular solutions of the Yang-Baxter equation of eight-vertex type. We find four independent solutions of which two are new. We were able to relate some AdS/CFT integrable models to our new models and in this way we could classify their integrable deformations. The $\mathrm{AdS}_3$ $R$-matrices we found correspond to the case of same chirality, while the $R$-matrices of opposite chirality are not regular and can instead be obtained up to some constants by requiring that they satisfy the Yang-Baxter equation \cite{new}. It is interesting that we can deform the two matrices with the same chirality independently. There are many new pressing open questions and future directions for research. 

First, it would be interesting to apply our method to a wider range of physical systems. In particular the case of a four-dimensional local Hilbert space is of interest as it would contain the Hubbard model and generalizations thereof. In this way deformations of the $\mathrm{AdS}_5$ superstring could also be studied. We plan to address some of these issues in upcoming work \cite{new}.

Second, it would be important to study and understand the physical and mathematical properties of the new solutions of the Yang-Baxter equation that we  derived. For instance, one obvious direction would be computing the spectrum of the 8-vertex B model by performing the Bethe Ansatz. Similarly, it would be interesting to find out if there is a quantum algebra and associated Yangian symmetry $ \mathcal{Y} $ underlying our new solutions. It is also unclear if there are 1+1 dimensional integrable field theories whose two-body scattering matrix corresponds to our new solutions

Third, we need to find an interpretation for the deformations for the holographic integrable models. In particular, the meaning of the deformation parameters on both the String and CFT side should be worked out. Understanding the functional (infinite dimensional) deformation of the $\mathrm{AdS}_3$ model should also be very fascinating.

Fourth, our method raises further questions regarding the general structure of integrable models. So far, imposing \eqref{eq:IntCond} is actually sufficient. This seems to support an old conjecture for integrability \cite{Grabowski}. However, it is unclear why this is the case and attempts at proving it have failed. Moreover, we do not impose braiding unitarity, 
$R_{12}(u,v) R_{21}(v,u) \sim 1$,
but all our solutions satisfy it nevertheless.

Lastly, it would be interesting to consider long-range deformations of our models \cite{Bargheer:2008jt,Bargheer:2009xy, BFdLL2013integrable}. Such deformations of spin chains correspond to loop corrections in the AdS/CFT correspondence.

\smallskip



\begin{acknowledgments}
\paragraph{Acknowledgements.}

We would like to thank S. Ekhammar, V. Korepin,  V. Kazakov, O. Ohlsson Sax, L. Takhtadzhan and A. Torrielli for discussions. MdL was supported by SFI, the Royal Society and the EPSRC for funding under grants UF160578, RGF$\backslash$EA$\backslash$181011, RGF$\backslash$EA$\backslash$180167 and 18/EPSRC/3590. 
C.P. is supported by the grant RGF$\backslash$EA$\backslash$181011. 
A.P. is supported by the grant RGF$\backslash$EA$\backslash$180167. 
A.L.R. is supported by the grant 18/EPSRC/3590. 
The work of P.R. is supported in part by a Nordita Visiting PhD Fellowship and by SFI and the Royal Society grant UF160578.

\end{acknowledgments}

\bibliography{lettershortened}

\end{document}